\documentclass{emulateapj}
\usepackage[varg]{txfonts}
\usepackage{natbib,twoopt}
\bibpunct{(}{)}{;}{a}{}{,}
\pdfoutput=1

\shorttitle{BSS bimodality: a Brownian motion model}
\shortauthors{Pasquato et al.}

\begin{document}
\title{Blue Straggler bimodality: a Brownian motion model}

\author{Mario Pasquato}
\affil{INAF, Osservatorio Astronomico di Padova, vicolo dell'Osservatorio 5, 35122 Padova, Italy}

\author{Paolo Miocchi}
\affil{Istituto dei Sistemi Complessi, CNR, P.le A.~Moro 2, 00185 - Roma (Italy)}

\author{Suk-Jin Yoon}
\affil{Department of Astronomy \& Center for Galaxy Evolution Research, Yonsei University, Seoul 120-749, Republic of Korea}

\begin{abstract}
The shape of the radial distribution of Blue Straggler Stars (BSS), when normalized to a reference population
of Horizontal Branch (HB) stars, has been found to be a powerful indicator of the dynamical evolution reached
by a Globular Cluster (GC).
In particular, observations suggest that the BSS distribution bimodality is modulated by the dynamical age of
the host GC, with dynamically unrelaxed GCs showing a flat BSS distribution, and more relaxed GCs showing a
minimum at a radius that increases for increasing dynamical age, resulting in a natural ``dynamical clock''.
While direct N-body simulations are able to reproduce the general trend, thus supporting its dynamical origin,
the migration of the minimum of the distribution appears to be noisy and not well defined.
Here we show that a simple unidimensional model based on dynamical friction (drift) and Brownian motion (diffusion) correctly
reproduces the qualitative motion of the minimum, without adjustable parameters except for the BSS to HB stars
mass-ratio. Differential dynamical friction effects combine with diffusion in creating a bimodality in the BSS
distribution and determining its evolution, driving the migration of the minimum to larger radii over time.
The diffusion coefficient is strongly constrained by the need to reproduce the migratory behaviour of the minimum,
and the radial dependence of diffusion set by fundamental physical arguments automatically satisfies
this constraint. Therefore, our model appears to capture the fluctuation-dissipation dynamics that underpins the
dynamical clock.
\end{abstract}

\section{Introduction}
In Globular Cluster (GC) color-magnitude diagrams, Blue Straggler Stars (BSS) are located on an extension of the main-sequence, at magnitudes brighter than the turnoff \citep[][]{1953AJ.....58...61S}. Spectroscopic and photometric evidence shows that BSS are more massive than main-sequence stars, with a typical mass of $\approx 1.2 M_\odot$ \citep[][]{1997ApJ...489L..59S, 1998ApJ...507..818G, 2005ApJ...632..894D, 2006ApJ...647L..53F, 2007ApJ...663.1040L, 2014ApJ...783...34F}. Two channels for the formation of BSSs were proposed: mass transfer in binary systems \citep[][]{1964MNRAS.128..147M, 1976ApJ...209..734Z} and direct stellar collisions \citep[][]{1976ApL....17...87H}. The two formation mechanisms may be working simultaneously with their relative importance determined by the environment \citep[see][and citations therein]{2009Natur.462.1028F}.

The radial distribution of BSS, normalized to a reference population (as Red Giants or Horizontal Branch stars) following the prescriptions
by \cite{1993AJ....106.2324F}, has been found to be diverse. In fact, it has been found to be bimodal (with a high peak in the
cluster center, a dip at intermediate radii and a rising branch in the
external regions) in the majority of the investigated GCs  
\citep[][]{1993AJ....106.2324F, 1997A&A...324..915F, 2004ApJ...603..127F,
2007ApJ...670.1065L,2007ApJ...663..267L}.
 In a few GCs the BSS radial distribution shows a
central peak, followed by a flat behavior out to the most external
regions (see NGC 1904, \citealt[][]{2007ApJ...663.1040L}). In other
systems ($\omega\,$Centauri and NGC 2419; \citealt{2006ApJ...638..433F, 2008ApJ...681..311D}) the distribution is
completely flat and not even a central peak is detected.
 
On the basis of these observational facts,
\citep[][F2012 in the following]{2012Natur.492..393F} proposed that the normalized BSS distribution can be used as a dynamical clock to measure the dynamical age of GCs. F2012 used the clock to classify a sample of $21$ Galactic GCs into families of different dynamical age based on the shape of their BSS radial distribution:
Family I with flat distribution as dynamically young clusters, Family II with bimodal distribution as dynamically intermediate-age and Family III with single peak distribution as dynamically old clusters. 

\cite{2004ApJ...605L..29M, 2006MNRAS.373..361M, 2007MNRAS.380.1127M, 2009MNRAS.396.1771M} have interpreted the formation of the bimodality as an effect of dynamical friction, and have reproduced it with Montecarlo simulations, both in GCs and dwarf galaxies.
\cite{2015ApJ...799...44M} have reproduced the formation of the bimodality with direct N-body simulations, with results that are qualitatively compatible with observations. However,  the description of the evolution in time of the bimodality in the simulations remains quite noisy. The \cite{2015ApJ...799...44M} simulations included only BSS formed through the mass-transfer channel and modeled both the BSS and their progenitors as single massive particles evolving in a background of lighter stars. This does not exclude the possibility that collisional BSS play a role in the bimodality, but it shows that the mass-segregation dynamics of mass-transfer BSS driven by dynamical friction is able to reproduce the bimodal distribution and its characteristic temporal evolution. Further simulations \citep[][]{2014ApJ...795..169A} showed that the timescale over which dynamical friction acts on stars is monotonically increasing as a function of radius even when considering the case of a background of stars with mass spectrum \citep[][]{2010AIPC.1242..117C} within the GC. The latter suggests that a progressive erosion of the BSS population at progressively larger radii is the most economical explanation of the shape of the observed radial distribution.

In this paper we attempt to capture the mechanisms underpinning the dynamical clock by introducing a self-consistent, physically motivated minimal model inspired by the progressive erosion picture. The point of the model presented in this paper is to clarify which mechanisms are sufficient for driving the formation and motion of the minimum, so we minimize the number of dynamical ingredients included.

As a result we understand that mass-segregation induced by dynamical friction and some sort of effective diffusion are the only ingredients needed to obtain a working dynamical clock, as long as the intensity of diffusion at least approximately respects the constraint of a fluctuation-dissipation relation such as the Einstein-Smoluchowski condition \citep[][]{2005AnP...517S.182E, 1906AnP...326..756V}.

\section{The model}
We model the motion of both the BSS stars and the reference HB stars in a background of lighter stars, distributed as a \cite{1911MNRAS..71..460P} model, as a one-dimensional Brownian motion. The actual Plummer potential in the three-dimensional problem is central, so the motion of any star - neglecting two-body encounters - takes place in a plane and is a rosetta orbit $\rho = \rho(t)$ confined between a minimum distance $r_\mathtt{min}$ and a maximum distance $r_\mathtt{max}$ from the cluster center.
A good measure of the scale radial position of the star, i.e. the scale size of the orbit, is the time-averaged distance from the center:
\begin{equation}
r = \frac{\int_{r_\mathtt{min}}^{r_\mathtt{max}} \rho \frac{d\rho}{\dot{\rho}(\rho)}}{\int_{r_\mathtt{min}}^{r_\mathtt{max}} \frac{d\rho}{\dot{\rho}(\rho)}}
\label{semiclassical}
\end{equation}
In the following we are not interested in working out the actual stochastic differential equations that would result in an accurate description of even the simplified three-dimensional problem of a star's motion in the Plummer potential plus two-body encounters. Instead we assume that the scale radius of stellar orbits as calculated in Eq.~\ref{semiclassical} evolves due to two-body encounters according to the following one-dimensional Langevin equation, that holds approximately for quasi-linear orbits, i.e. at low angular momentum:
\begin{equation}
\label{Langevin}
m \ddot{r} = - \alpha \dot{r} + m \Gamma(r) + \mathbf{\eta}(t)
\end{equation}
where $m$ is the star mass, $\alpha$ is a damping coefficient that we will discuss in the following, $\Gamma(r)$ is an effective gravitational field, and $\mathbf{\eta}(t)$ is a stochastic force term.
This is a strong simplification, but with an adequate choice of $\Gamma(r)$, the overdamped motion of this system matches the ``bead-in-honey'' dynamics of the qualitative picture underlying the BSS radial distribution evolution in e.g. \cite{2006MNRAS.373..361M, 2012Natur.492..393F}. The point of this exercise is to show that the shape of the BSS radial distribution is influenced by exactly two ingredients, namely orbital decay due to dynamical friction $- \alpha \dot{r} + m \Gamma(r)$ and diffusion $\mathbf{\eta}(t)$.
The Plummer gravitational field itself is an obvious choice for $\Gamma(r)$:
\begin{equation}
\Gamma(r) = g(r) = - \frac{G M(r)}{r^2} = - \frac{G M}{R^2} \frac{r/R}{(1 + r^2/R^2)^{3/2}}
\end{equation}
and $\eta$ is a random force that is normally distributed with zero mean and no memory (its autocorrelation is a delta function). As mentioned above we are interested in the overdamped motions where $\ddot{r} = 0$ and the star's orbit scale radius acquires a drift velocity
\begin{equation}
\label{drift}
v_d(r) = \frac{m g(r)}{\alpha} = - \frac{G M m}{\alpha R^2} \frac{{r}/{R}}{{\left( 1 + {r^2}/{R^2} \right) }^{3/2}}
\end{equation}
where in the latter equality we introduced the Plummer scale parameters $M$ and $R$ \citep[see][]{1987degc.book.....S} and temporarily ignored the random component of Eq.~\ref{Langevin}.
This corresponds to a dynamical friction timescale that increases with radius as
\begin{equation}
\label{dft}
\tau(r) = - \frac{r}{v_d(r)} =  \frac{\alpha R^3}{G M m} {\left( 1 + \frac{r^2}{R^2} \right) }^{3/2}
\end{equation} 
which embodies our expectation that the evolution is faster for more massive stars (increasing the mass decreases the timescale) and that faraway stars evolve more slowly ($\tau$ is monotonic in radius).
The parameter $\alpha$ can be fixed by requiring stars that reside in the center of the cluster (at $r = 0$) to have a dynamical friction timescale compatible with Eq.~1 of \cite{2006MNRAS.373..361M}:
\begin{equation}
\label{Mapellicondition}
\tau(0) = \frac{\alpha R^3}{G M m} = \frac{3}{4 \log{\Lambda} G^2 \sqrt{2 \pi}} \frac{\sigma^3(0)}{m \rho(0)}
\end{equation}
where $\sigma(0)$ and $\rho(0)$ are the velocity dispersion and density at the center of our Plummer model, and $\log{\Lambda}$ is the Coulomb logarithm of the system.
Equation \ref{Mapellicondition} implies that $\alpha$ is independent on the mass of the star undergoing dynamical friction. It is
\begin{equation}
\label{defalpha}
\alpha = \frac{\sqrt{\pi}}{4 \log{\Lambda}} \sqrt{\frac{G M}{R^3}} M
\end{equation}
which is a numeric factor times the characteristic mass $M$ of the Plummer model over its crossing time $T_c$. Incidentally, the damping timescale in Eq.~\ref{Langevin} is
\begin{equation}
\frac{m}{\alpha} \approx \frac{m}{M} T_c \ll T_c
\end{equation}
so our result is still within the regime compatible with the initial overdamping assumption. By using Eq.~\ref{defalpha}, we can rewrite Eq.~\ref{dft} in the form
\begin{equation}
\label{physicaltau}
\tau(r) = \frac{\sqrt{\pi}}{4 \log{\Lambda}} \frac{N}{\sqrt{{G M}/{R^3}}} \frac{\langle m \rangle}{m}  {\left( 1 + \frac{r^2}{R^2} \right) }^{3/2} \propto T_c \frac{\langle m \rangle}{m}  {\left( 1 + \frac{r^2}{R^2} \right) }^{3/2}
\end{equation}
where $\langle m \rangle = M/N$ is the mean mass of the stars in the system, and $T_c$ is the core relaxation time.
The Coulomb logarithm $\log{\Lambda}$ determines the physical scaling of our model (the way to convert its predictions to physical units of parsecs, years, and solar masses) through Eq.~\ref{physicaltau}.

It should be noted that Eq.~1 of \cite{2006MNRAS.373..361M} enters our model only to determine the numerical value of $\alpha$ (essentially by dimensional considerations) but it describes a slightly different dynamics than ours. To have a dependence of $\tau(r)$ that reflects Eq.~1 of \cite{2006MNRAS.373..361M} we would need to have a $7/4$ exponent in Eq.~\ref{dft} in place of $3/2$. 

The motion of stars due to the drift velocity in Eq.~\ref{dft} is simple and can be solved analytically by solving $\dot{r} = v_d(r)$ with the relevant initial conditions for a given star, i.e. $r = r_0 = r(t = 0)$, obtaining
\begin{equation}
\label{anal}
\int_{r_0}^r \frac{d \xi}{v_d(\xi)} = t
\end{equation}
which can be solved for $r$ giving $r = r(r_0, t)$. However, it turns out that it is simpler to proceed numerically, which also allows us to seamlessy introduce the effects of Brownian motion, i.e. diffusion. So we consider the evolution of a star's position $r$ over a timestep $\delta t$:
\begin{equation}
\label{numevol}
r^\prime = r + v_d(r) \delta t + \sqrt{D(r) \delta t} \hat{\eta}
\end{equation}
where $r^\prime$ is the new position of the star after the evolution step, $v_d(r) \delta t$ is responsible for the evolution under the drift velocity due to dynamical friction, and the last term represents the effect of the random motion. In particular, $\hat{\eta}$ is a random number extracted from a normal distribution with mean zero and unit variance, and $D(r)$ has the dimensions of a diffusion coefficient. This numerical evolution step, repeated over a population of stars, results in a collective behaviour that is described by Fick's law of diffusion:
\begin{equation}
\label{Fick}
J = J_d - D \frac{\partial n}{\partial r}
\end{equation}
where $J$ is the one-dimensional flux of stars that move due to the combination of systematic drift motion ($J_d$) and diffusion (represented by the second term), and $n$ is their number density. Equation \ref{Fick} shows that diffusion prevents the formation of large density gradients, smoothing out sharp peaks, because the diffusive flux is inversely proportional to the derivative of density. The larger $D$, the stronger this effect. However, $D$ cannot be selected arbitrarily, because the dynamical friction determining the drift velocity and diffusion both arise from the same underlying phenomenon, random uncorrelated collisions with lighter stars. Therefore $D(r)$ is related to $v_d(r)$ according to the Einstein-Smoluchowski relation \citep[see e.g.][]{1966RPPh...29..255K} that connects the diffusion coefficient to mobility $\mu$:
\begin{equation}
D = \mu k T
\end{equation}
where $k$ is the Boltzmann constant and $T$ is the temperature. This holds at thermal equilibrium in a gas or in a solution (in the case of the original Brownian motion), but we will substitute here $k T$ with the typical kinetic energy of stars as a function of radius, i.e.
\begin{equation}
\label{smolu}
D(r) = \mu m \sigma^2(r)
\end{equation}
where
\begin{equation}
\sigma^2(r) = \frac{GM}{2 R} \frac{1}{\sqrt{1 + r^2/R^2}}
\end{equation}
is the squared velocity dispersion of stars in our Plummer model. The mobility $\mu$ is defined as
\begin{equation}
\label{mobility}
\mu(r) = \frac{v_d(r)}{m g(r)} = \frac{1}{\alpha}
\end{equation}
so
\begin{equation}
\label{smoluunpack}
D(r) = \frac{GMm}{2 R \alpha} \frac{1}{\sqrt{1 + r^2/R^2}}
\end{equation}
and we can finally use Eq.~\ref{numevol} to evolve the system in a self-consistent way.

Additionally, Eq.~\ref{numevol} can be rewritten as
\begin{equation}
\label{numevol2}
r^\prime = r + \chi g(r) + \sqrt{\chi} \sigma(r) \hat{\eta}
\end{equation}
where the constant
\begin{equation}
\chi = \frac{m \delta t}{\alpha}
\end{equation}
has the dimensions of time squared. The evolution depends only on this combination of mass and time, so that more massive stars have the same evolution as less massive stars, but speeded up by a factor proportional to the mass ratio. In the following we will evolve two species of stars, BSS progenitors and reference stars such as HBs, both with the same timestep $\delta t$, so the ratio
\begin{equation}
\frac{\chi_{\mathtt{BSS}}}{\chi_{\mathtt{HB}}} = \frac{m_{\mathtt{BSS}}}{m_{\mathtt{HB}}}
\end{equation}
determines whose evolution is faster. We set
\begin{equation}
\frac{m_{\mathtt{BSS}}}{m_{\mathtt{HB}}} = \frac{3}{2}
\end{equation}
In this way the BSS evolve $1.5$ times as fast as the $HB$, as expected of $1.2 M_\odot$ stars with respect to $0.8 M_\odot$ stars.

We used Eq.~\ref{numevol2} to evolve $5 \cdot 10^{4}$ BSS and $5 \cdot 10^{4}$ HBs for $20000$ timesteps until $t = 80 \tau(0)$ with $\tau$ as defined by Eq.~\ref{dft}, which in physical units is:
\begin{equation}
\tau(0) = \frac{\sqrt{\pi}}{4 \log{\Lambda}} \frac{R}{\sqrt{GM/R}} \frac{M}{m}
\end{equation}
Stars have initially the same Plummer distribution as the one used to calculate the effects of dynamical friction, so there is initially no mass-segregation. We force stars to stay within the region $0 < r < 8R$ by placing stars that move to $r > 8 R$ due to the evolution back to $8 R$, and by flipping the sign of $r$ if it becomes negative (the boundary at $r = 0$ is reflective).
We divide the stars in $20$ equally spaced radial bins and calculate the ratios between BSS and HB in each bin. At the beginning $\mathtt{BSS}/\mathtt{HB}$ is constant and equals $1$ at every radius, within the Poisson counting error. In the presentation of our results we adopt units so that $GM = 1$, $R = 1$, and $\tau(0) = 1$.

\section{Results and interpretation}
Qualitatively, the different drift velocity of BSS and HB stars, due to their different mass, results over time in an higher concentration of BSS in the center of the cluster, so that starting with a $\mathtt{BSS}/\mathtt{HB}$ ratio constant and equal to unity at every radius, $\mathtt{BSS}/\mathtt{HB}$ evolves to form a peak in the cluster center. If nothing else intervenes, dynamical friction eventually drives all stars to $r = 0$, resulting in an unphysical outcome: while BSS are more concentrated than HBs in relaxed clusters, their ratio is still limited to $\approx 3$, as shown in F2012. In our model, though, diffusion counteracts this tendency of dynamical friction to pile up all the heavy stars in the cluster center. This is expected in any self consistent dynamical model, given that dynamical friction and diffusion are two sides of the same phenomenon. It is the combination of dynamical friction and diffusion that shapes $\mathtt{BSS}/\mathtt{HB}$ resulting in a bimodal distribution with a minimum that moves outwards over time, i.e. in a dynamical clock. The model presented in this paper represents a theoretical understanding of the mechanism that moves the hands of the clock.

\begin{figure}
\includegraphics[width=0.95\columnwidth]{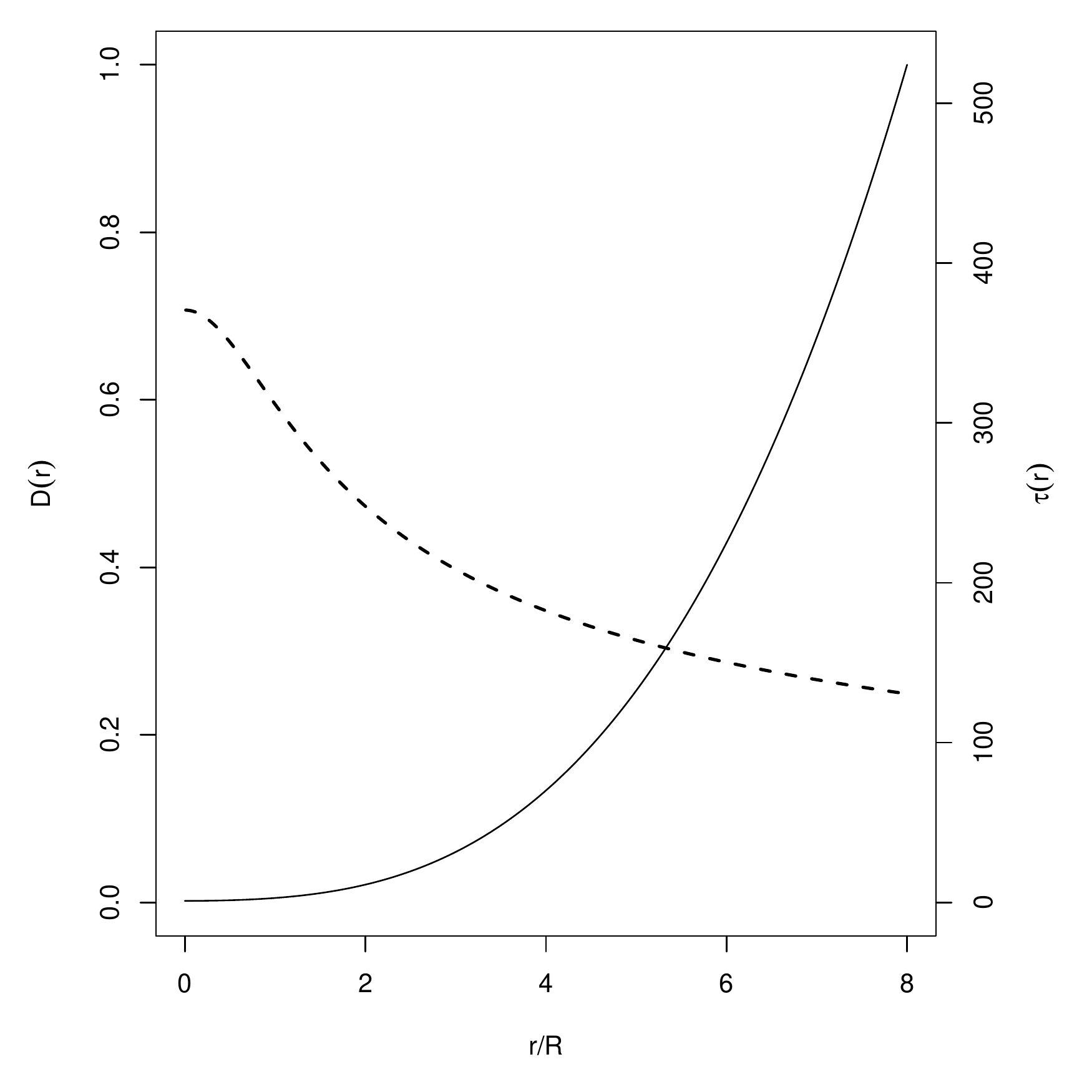}
\caption{Diffusion coefficient $D(r)$ shown as a function of radius (dashed thick line, scale on the left vertical axis) together with the dynamical friction characteristic timescale $\tau(r)$ (solid thin line, scale on the right axis). The dynamical friction timescale depends strongly on radius, much less so the diffusion coefficient.\label{dridi}}
\end{figure}

This can be better understood by looking at Fig.~\ref{dridi}, where the diffusion coefficient $D(r)$ and the dynamical friction timescale $\tau(r)$ are plotted as a function of $r/R$ for our Plummer model. In the outskirts, stars drift very slowly towards the center (the timescale for dynamical friction is long) and also are relatively unaffected by diffusion (the diffusion coefficient is small), so the $\mathtt{BSS}/\mathtt{HB}$ ratio stays around its initial value. Stars that are located at intermediate radii are still relatively unaffected by diffusion but have a much shorter timescale for dynamical friction, and fall towards the center. This affects BSS more than HB, because they are heavier. Thus the $\mathtt{BSS}/\mathtt{HB}$ ratio drops at intermediate radii. In the center BSS stars pile up, so that the $\mathtt{BSS}/\mathtt{HB}$ ratio increases, but the amount of this increase is limited by diffusion, that is stronger in the center and spreads out peaks in the density distribution. 

\begin{figure}
\includegraphics[width=0.95\columnwidth]{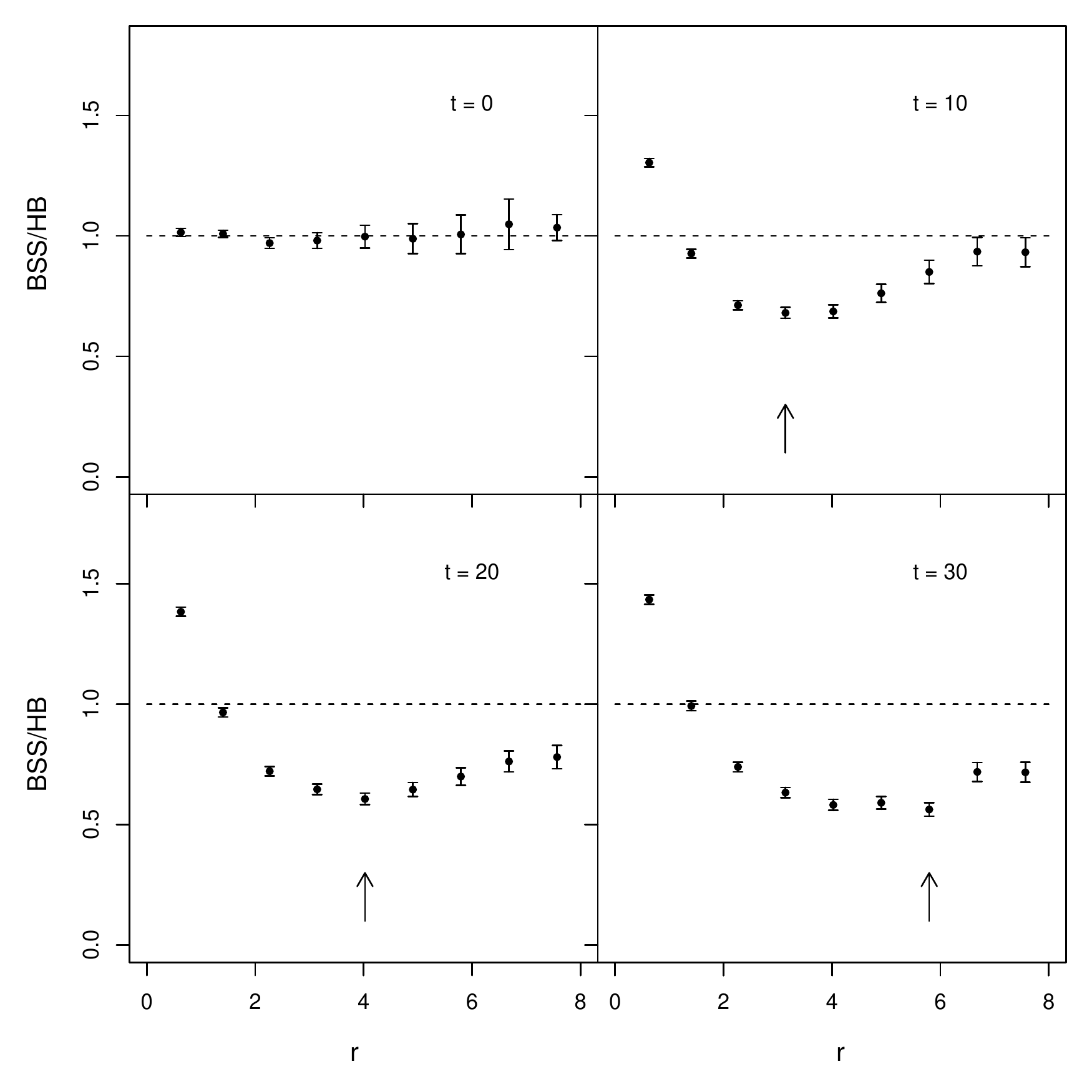}
\caption{Temporal evolution of the BSS/HB ratio in the model. The panels show the BSS/HB ratio in radial bins (3D radius) at increasing times (left to right, top to bottom). The minimum of the BSS/HB ratio is indicated by an arrow. The dashed lines corresponds to the initial value of BSS/BH $= 1$, which holds in every bin within the errors. Error bars represent one sigma Poisson counting error. The minimum forms and moves to larger radii as as the cluster ages.\label{tevo}}
\end{figure}

\begin{figure}
\includegraphics[width=0.95\columnwidth]{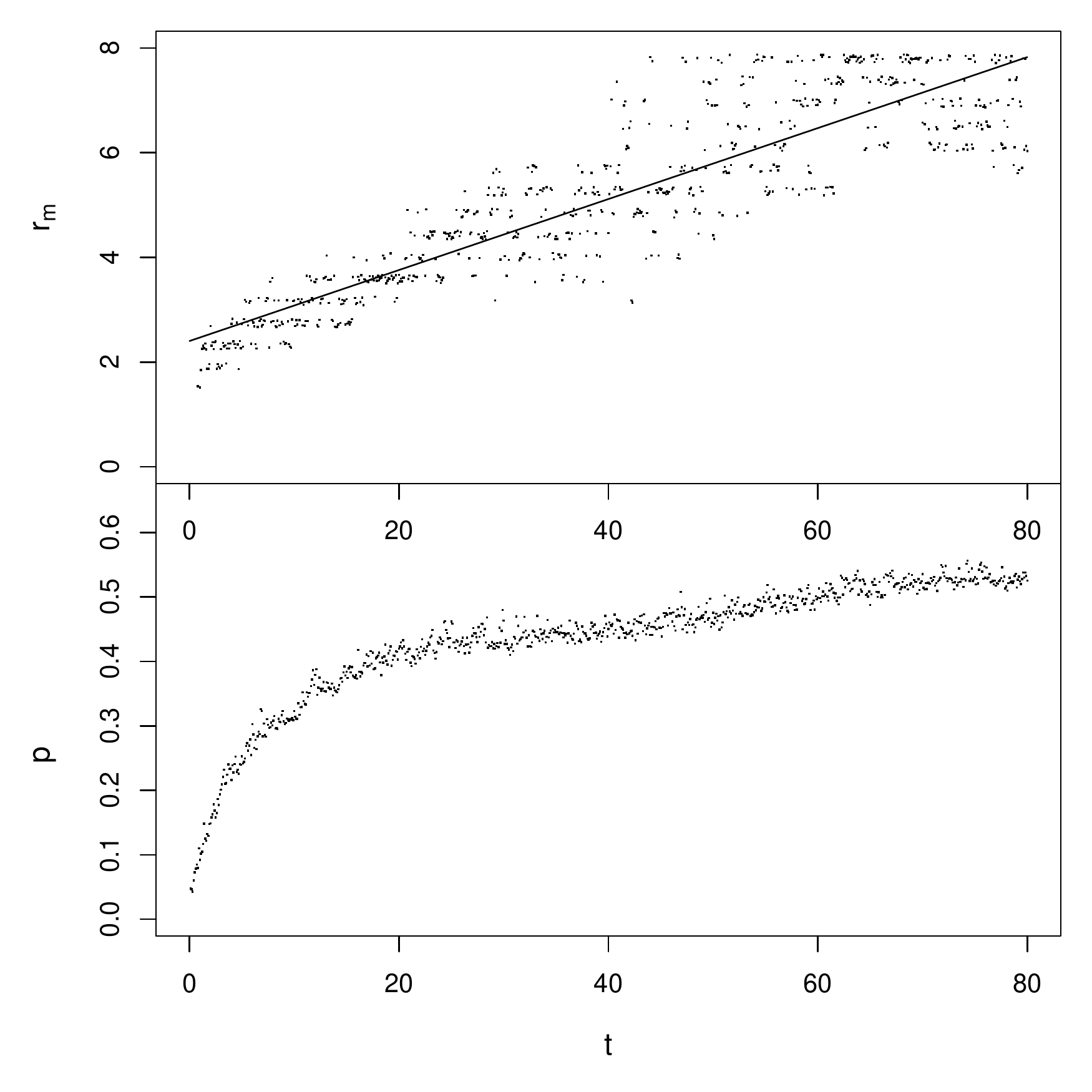}
\caption{Upper panel: radial position $r_m$ of the minimum of BSS/HB as a function of time in a typical run. Only minima that differ from $1$ at a three-sigma level have been included in the plot. The minimum moves outwards with time. A linear best fit is superimposed (black solid line). To improve visualization some random jittering was added to the ordinates to break ties. Lower panel: depth $p$ of the minimum as a function of time. The depth of the minimum is defined as $1 - \mathtt{BSS}/\mathtt{HB}$ at $r_m$. \label{mevo}}
\end{figure}

Figure \ref{tevo} shows the temporal evolution of the $\mathtt{BSS}/\mathtt{HB}$ ratio in our model. At increasing times (time increasing in the panels from left to right, top to bottom) the minimum moves towards larger radii in a linear fashion. Fig.~\ref{mevo} shows the evolution of the position and the depth of the minimum of the $\mathtt{BSS}/\mathtt{HB}$ ratio. The position of the minimum evolves linearly with time.

\subsection{Constant diffusion coefficient models: a fine-tuning problem}
The importance of Eq.~\ref{smolu} in revealing the dynamics of the system is better understood if we build models where dynamical friction acts as usual, but the diffusion coefficient is set to a constant independent of radius. If we were to ignore the fundamental physical constraint that ties dynamical friction and diffusion, the most natural way to proceed would be to model dynamical friction alone. When this results in an unphysical outcome (diverging density and $\mathtt{BSS}/\mathtt{HB}$ peak in the center) the next step is to add diffusion to limit the effects of dynamical friction. However, lacking guidance from Eq.~\ref{smolu}, it is not clear exactly \emph{how much diffusion} is needed. While in this setting the naive assumption of a radially constant diffusion coefficient seems appropriate, there is a priori no reason, bar dimensional considerations, to choose a specific value over a different one.

We thus build constant $D$ models with different values of $D$ and observe the system's behaviour in terms of formation and motion of the minimum. These are the two ingredients of the dynamical clock: the minimum has to form, i.e. the normalized BSS distribution needs to become appreciably bimodal at some point in time, and its minimum needs to move to larger radii over time for the clock to work. It turns out that too large a $D$ (too much diffusion) prevents the formation of the minimum, because by smoothing out strong gradients in the density distribution of stars, it spreads out the minimum to the point that it becomes unobservable. On the other hand if $D$ is too small (not enough diffusion) dynamical friction is able to create a very sharp and significant minimum very quickly, but the minimum does not move to larger radii as time goes by. It is the smoothing out of the central $\mathtt{BSS}/\mathtt{HB}$ peak that pushes the minimum outwards, so diffusion and dynamical friction have to work in a concerted way to produce a minimum and make sure it shifts to larger radii over time. In the setting of arbitraily chosen constant diffusion coefficients this is effectively a fine-tuning problem. Including a physically motivated dependence of $D$ on cluster radius solves this problem in a remarkably elegant way. This is a demonstration that, no matter how complex the phenomena underlying the formation and motion of the minimum in actual clusters and N-body simulations, the dynamical friction plus diffusion model captures a profound aspect of the underlying fluctuation/dissipation dynamics of the system.

\begin{figure}
\includegraphics[width=0.95\columnwidth]{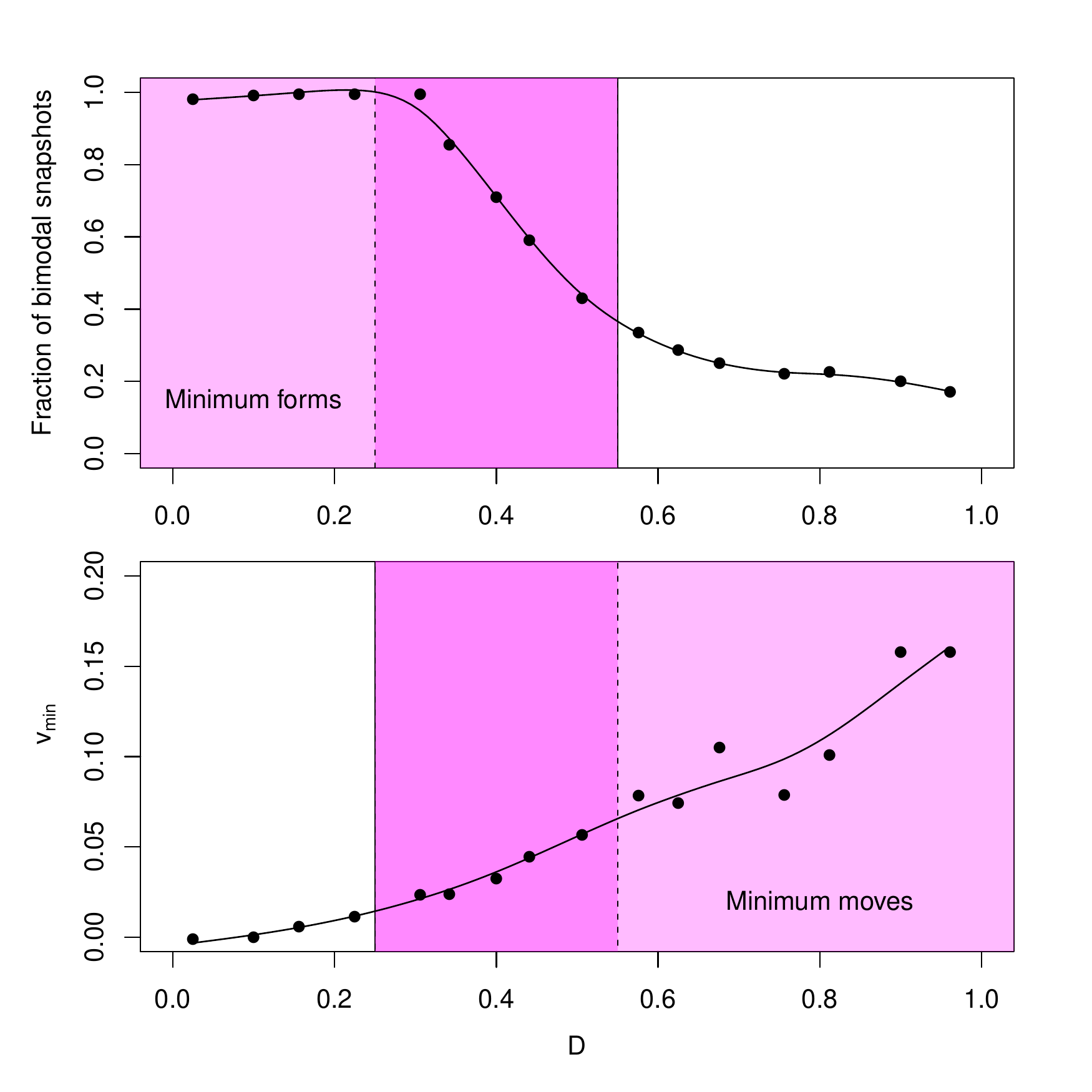}
\caption{Effects of different constant diffusion coefficients $D$ on the formation and motion of the minimum. The upper panel shows the fraction of snapshots where a significant minimum forms as a function of the diffusion coefficient of a given run (filled circles), with a cubic spline smoothing line superimposed (solid line). To the left of the vertical solid line that delimits the pink area, at least one third of the snapshots is bimodal, so we can say that a minimum forms. The lower panel shows the speed $v_{min}$ of the outward motion of the minimum as a function of $D$ (filled circles, smoothed solid line). To the right of the vertical solid line the minimum of the $\mathtt{BSS}/\mathtt{HB}$ ratio moves outwards at an appreciaby within the cluster lifetime (pink area). The dark pink area in both panels represents the region where both conditions are met, i.e. the minimum forms and moves outwards. If a constant value of $D$ is picked arbitrarily there is no a priori reason it would fall in the dark pink area, thus giving rise to a fine-tuning problem, which is solved by the physically motivated dependence of $D$ on radius as discussed in the previous sections.\label{cevo}}
\end{figure}

Figure \ref{cevo} shows that large values of $D$ prevent the formation of a minimum (upper panel). The number fraction of snapshots from a constant-D model where the distribution is significantly bimodal, as defined by having a minimum that is significantly lower than the $\mathtt{BSS}/\mathtt{HB}$ in the outermost bin at three-sigma, is plotted as a function of $D$. We see a plateau at low diffusion coefficients where almost all snapshots are bimodal, and a sharp drop when $D$ is increased until only about $20\%$ of snapshots are bimodal. The lower panel shows the velocity of the outward motion of the minimum obtained by a linear fit of the position of the minimum as a function of time (only for the snapshots that are bimodal). Modulo some fluctuations the velocity increases monotonically with $D$ showing that more diffusion pushes the minimum to larger radii faster. In Fig.~\ref{cevo} we show in pink the areas where the minimum forms (upper panel) and where the minimum moves (lower panel). The $D$ values for which both happen (the dynamical clock works) correspond to the dark pink area, which unsurprisingly corresponds to the range of $D$ values spanned by $D = D(r)$ in the physically self-consistent model in Fig.~\ref{dridi}. In the setting of constant-D models, instead, there is no reason why $D$ would fall in the dark pink area, raising the fine-tuning problem discussed above. 

\section{Conclusions}
We presented a new simplified unidimensional model of the evolution of BSS stars in a GC based on dynamical friction and diffusion of stars in almost radial orbits in a static Plummer potential. Our model has no free parameters and minimal input physics, with a self-consistent treatment of diffusion which is coupled to dynamical friction via the Einstein-Smoluchowski relation.

Despite its simplicity, this model qualitatively reproduces the observed evolution of the normalized BSS radial distribution. In particular it correctly predicts the formation of a minimum at intermediate radii and its progressive outwards motion (dynamical clock). Our results are qualitatively compatible with the first studies of BSS bimodality based on Montecarlo simulations (that similarly included dynamical friction and diffusion) such as \cite{2004ApJ...605L..29M} and with direct N-body simulations \citep{2015ApJ...799...44M}, despite noise problems in the latter.
This is remarkable because a simplified (but admittedly unphysical) version of our models including only dynamical friction and not diffusion fails to reproduce the outwards motion of the minimum, and models with a constant, arbitrarily selected diffusion coefficient need fine tuning of the coefficient to reproduce both the minimum formation and motion. The fact that N-body models such as \cite{2015ApJ...799...44M}, which have a number of particles at least an order of magnitude lower than real clusters, are still in the right dynamical regime - i.e. are capable of generating a minimum and making it move outwards - is non trivial. It would be certainly extremely interesting to explore the parameter space (in terms of number of particles, initial phase space distribution, and mass and number ratios between particles representing BSS, RGB, and field stars) over which N-body simulations are able to consistently achieve a BSS minimum formation and motion in a statistically significant way.

Our results shed light on the mechanisms underlying the dynamical clock proposed by \cite{2012Natur.492..393F}. It shows that, no matter what complex phenomena take place in real GCs and in N-body simulations, as long as these phenomena produce an effective dynamical friction and an effective diffusion that are at least approximately following the Einstein-Smoluchowski condition, they will result in a working dynamical clock. This does not logically imply, but it strongly suggests that our simple model captures the physics that makes the dynamical clock tick.

\section*{Acknowledgements}
This project has received funding from the European UnionÕs Horizon $2020$ research and innovation programme under the Marie Sk\l{}odowska-Curie grant agreement No. $664931$. Additionally, M.P. and S.-J.Y. acknowledge support from Mid-career Researcher Program (No. 2015-008049) through the National Research Foundation (NRF) of Korea. S.-J.Y. acknowledges support from Mid-career Researcher Program (No. 2015-008049) through the National Research Foundation (NRF) of Korea, from the Center for Galaxy Evolution Research (No. 2017R1A5A1070354) through the NRF of Korea,
and from the Yonsei--KASI joint research grant through the Korea Astronomy and Space Science Institute Research Fund 2017-2018. We thank Prof. Francesco R. Ferraro, Prof. Barbara Lanzoni, Prof. Enrico Vesperini, Prof. Michela Mapelli, and Dr. Stefano Pugnetti for very helpful comments and suggestions.

\bibliography{manuscript}
\bibliographystyle{aa}

\end{document}